\documentclass[twocolumn,prl,superscriptaddress,floatfix,preprintnumbers,letterpaper]{revtex4}

\usepackage{epsfig}
\usepackage{graphics}
\usepackage{graphicx}
\usepackage{amsmath,amssymb}
\usepackage{amsfonts}
\usepackage{bm}
\usepackage[usenames,dvipsnames]{color}
\usepackage{amssymb}
\usepackage{psfrag}
\usepackage{times}
\usepackage[varg]{txfonts}
\usepackage[colorlinks, pdfborder={0 0 0}]{hyperref}
\definecolor{LinkColor}{rgb}{0.75, 0, 0}
\definecolor{CiteColor}{rgb}{0, 0.5, 0.5}
\definecolor{UrlColor}{rgb}{0, 0, 0.75}
\hypersetup{linkcolor=LinkColor}
\hypersetup{citecolor=CiteColor}
\hypersetup{urlcolor=UrlColor}

\usepackage{ulem}
\begin{document}

%%%%%%%%%%%%%%%%%%%%%%%%%%%%%%%%%%%%%%% Some newcommands %%%%%%%%%%%%%%%%%%%%%%%%%%%%%%%%%%%%
\newcommand{\be}{\begin{equation}}
\newcommand{\ee}{\end{equation}}
\newcommand{\ber}{\begin{eqnarray}}
\newcommand{\eer}{\end{eqnarray}}
\newcommand{\bea}{\begin{eqnarray}}
\newcommand{\eea}{\end{eqnarray}}
\newcommand{\ie}{i.e.}
\newcommand{\dt}{{\rm d}t}
\newcommand{\df}{{\rm d}f}
\newcommand{\hc}{{\sf h}}
\newcommand{\fmerg}{f_{1}}
\newcommand{\fring}{f_{2}}
\newcommand{\fcut}{f_{3}}
\newcommand{\rmi}{{\rm i}}
\newcommand{\CC}{\mathcal{C}}

\newcommand{\LIGOCaltech}{LIGO Laboratory, California Institute of Technology, 
Pasadena, CA 91125, USA}
\newcommand{\TAPIR}{Theoretical Astrophysics, California Institute of Technology, 
Pasadena, CA 91125, USA}
\newcommand{\Cork}{Physics Department, University College Cork, Cork, Ireland}
\newcommand{\UIB}{Departament de F\'isica, Universitat de les Illes Balears, 
Crta. Valldemossa km 7.5, E-07122 Palma, Spain}
\newcommand{\Jena}{Theoretisch-Physikalisches Institut, Friedrich Schiller Universit\"at Jena,
Max-Wien-Platz 1, 07743~Jena, Germany}
\newcommand{\AEIGolm}{Max-Planck-Institut f\"ur Gravitationsphysik (Albert-Einstein-Institut), 
Am~M\"uhlenberg 1, 14476~Golm, Germany}
\newcommand{\Vienna}{Gravitational Physics, Faculty of Physics, University of Vienna, Boltzmanngasse
5, A-1090 Vienna, Austria}
\newcommand{\Cardiff}{School of Physics and Astronomy, Cardiff University, Queens Building, CF24 3AA, Cardiff, United Kingdom}
\newcommand{\GSFC}{NASA Goddard Space Flight Center, 8800 Greenbelt Rd, Greenbelt, MD 20771, USA}

%%%%%%%%%%%%%%%%%%%%%%%%%%%%%%%%%%%%%%%%%%%%%%%%%%%%%%%%%%%%%%%%%%%%%%%%%%%%%%%%%%%%%%%%%%%%%

%%%%%%%%%%%%%%%%%%%%%%%%%%%%%%%%%%%%%% Title page %%%%%%%%%%%%%%%%%%%%%%%%%%%%%%%%%%%%%%%%%%%
\title{Inspiral-merger-ringdown waveforms for black-hole binaries with non-precessing spins} 

\author{P.~Ajith}
\affiliation{\LIGOCaltech}
\affiliation{\TAPIR}
%\affiliation{\AEIHann}

\author{M.~Hannam}
\affiliation{\Vienna}
\affiliation{\Cardiff}

\author{S.~Husa}
\affiliation{\UIB}

\author{Y.~Chen}
\affiliation{\TAPIR}

\author{B.~Br\"ugmann}
\affiliation{\Jena}

\author{N.~Dorband}
\affiliation{\AEIGolm}

\author{D.~M\"uller}
\affiliation{\Jena}

\author{F. Ohme}
\affiliation{\AEIGolm}

\author{D.~Pollney}
\affiliation{\AEIGolm}
\affiliation{\UIB}

\author{C.~Reisswig}
\affiliation{\AEIGolm}
\affiliation{\TAPIR}

\author{L.~Santamar\'ia}
\affiliation{\AEIGolm}
\affiliation{\LIGOCaltech}

\author{J.~Seiler}
\affiliation{\AEIGolm}

%%%%%%%%%%%%%%%%%%%%%%%%%%%%%%%%%%%%%%%%%%%%%%%%%%%%%%%%%%%%%%%%%%%%%%%%%%%%%%%%%%%%%%%

%%%%%%%%%%%%%%%%%%%%%%%%%%%%%%%%%%%%% Abstract %%%%%%%%%%%%%%%%%%%%%%%%%%%%%%%%%%%%%%%%
\begin{abstract}
We present the first analytical inspiral-merger-ringdown gravitational waveforms 
from binary black holes (BBHs) with non-precessing spins, that is based on a description 
of the late-inspiral, merger and ringdown in full general relativity. By matching 
a post-Newtonian description of the inspiral to a set of numerical-relativity simulations,
we obtain a waveform family with a conveniently small number of 
physical parameters. These waveforms will allow us to detect a 
larger parameter space of BBH coalescence, including a considerable fraction of precessing
binaries in the comparable-mass regime, thus significantly improving the expected detection 
rates.
\end{abstract}
%%%%%%%%%%%%%%%%%%%%%%%%%%%%%%%%%%%%%%%%%%%%%%%%%%%%%%%%%%%%%%%%%%%%%%%%%%%%%%%%%%%%%%%%

%%%%%%%%%%%%%%%%%%%%%%%%%%%%%%%%%%%% Preprint numbers %%%%%%%%%%%%%%%%%%%%%%%%%%%%%%%%%%
\preprint{LIGO-P0900085-v5}
%%%%%%%%%%%%%%%%%%%%%%%%%%%%%%%%%%%%%%%%%%%%%%%%%%%%%%%%%%%%%%%%%%%%%%%%%%%%%%%%%%%%%%%%

\maketitle

%%%%%%%%%%%%%%%%%%%%%%%%%%%%%%%%%%%% Introduction %%%%%%%%%%%%%%%%%%%%%%%%%%%%%%%%%%%%%%
Coalescing black-hole (BH) binaries are among the most promising
candidate sources for the first detection of gravitational
waves (GWs). Such observations will lead to precision tests of 
general relativity as well as provide a
wealth of information relevant to fundamental physics, astrophysics, 
and cosmology. Computation of the expected waveforms from these sources 
is a key goal in current research in gravitation. 

While the \emph{inspiral} and \emph{ring-down} stages of the BH coalescence 
are well-modeled by perturbative techniques, an accurate description of 
the \emph{merger} requires numerical solutions of Einstein's equations. 
Although performing numerical simulations densely sampling the entire parameter space
of BH coalescence is computationally prohibitive, waveform templates modeling all 
the three stages can now be constructed by combining analytical- and 
numerical- relativity results, dramatically improving the sensitivity of searches for 
GWs from BH binaries, and the accuracy of estimating the source parameters~\cite{Ajith:2007kx,
Ajith:2009fz,Buonanno:2007pf,Damour:2009kr}. To date, inspiral-merger-ringdown 
(IMR) templates have been computed only for nonspinning BH binaries~\cite{Ajith:2007qp,
Ajith:2007kx,Ajith:2007xh,Buonanno:2007pf,Damour:2009kr}. However, 
most BHs in nature are expected to be 
spinning~\cite{Volonteri:2004cf}, which necessitates the inclusion of 
spinning-binary waveforms in GW searches. But, spin adds six parameters (three components 
for each BH), and each additional parameter in a search template bank leads to a higher 
signal-to-noise-ratio (SNR) threshold for a confident detection~\cite{VanDenBroeck:2009gd}. 
Also, this requires accurate numerical simulations across this large parameter 
space, which are not yet available. Moreover, implementing a search covering 
the full spin parameter space has proven to be difficult. 

In this letter, we present an IMR waveform family modeling the dominant harmonic of 
binaries with non-precessing spins, i.e., spins (anti-)aligned with the orbital 
angular momentum. Aligned-spin binaries are an astrophysically interesting population 
as such systems are expected from isolated binary evolution and in gas-rich 
mergers~\cite{Kalogera:2004pr,2007ApJ...661L.147B}. Non-precessing binaries also exhibit interesting 
strong-gravity effects like the ``orbital hang-up''~\cite{Campanelli:2006uy} and ``spin 
flips''~\cite{Buonanno:2007sv}. We make use of the degeneracies in the physical parameters 
to parametrize our waveform family by only the total mass $M \equiv m_1+m_2$ of the binary, 
the symmetric mass ratio $\eta \equiv m_1m_2/M^2$, and a \emph{single} spin parameter 
$\chi \equiv (1+\delta) \, \chi_1/2 + (1-\delta)\,\chi_2/2$, where $\delta \equiv (m_1-m_2)/M$ 
and $\chi_i \equiv S_i/m_i^2$, $S_i$ being the spin angular momentum of the $i$th BH. 
The last feature is motivated by the observation that the 
leading spin-orbit-coupling term in post-Newtonian (PN) waveforms is dominated by this 
parameter. We also show that this waveform family is able to capture a significant 
fraction of precessing binaries in the comparable-mass regime, providing 
an efficient and feasible way of searching for these systems~\footnote{The 
reason is that the (spin-dependent) phase evolution is primarily governed by the spin-orbit coupling, 
determined by the spin components along the angular momentum.}. 

\paragraph{Numerical simulations.---}
Binary BH (BBH) waveforms covering at least eight wave cycles before merger
were produced by solving Einstein equations numerically, as
written in the ``moving-puncture''  3+1 formulation~\cite{Campanelli:2005dd,
Baker:2005vv}. The numerical solutions were calculated with the 
{\tt BAM}~\cite{Brugmann:2008zz,Husa:2007hp}, {\tt CCATIE}~\cite{Pollney:2007ss} 
and {\tt LLAMA}~\cite{Pollney:2009yz} codes. Initial momenta were chosen to 
give low-eccentricity inspiral, using either an extension of the method described 
in~\cite{Husa:2007rh}, or the quasicircular formula used in~\cite{Brugmann:2007zj}.
GWs were extracted at $R_{ex} = 90M$ with {\tt BAM}, $R_{ex} = 160M$ with {\tt CCATIE} 
and at future null infinity with {\tt LLAMA}, using procedures discussed 
in~\cite{Brugmann:2008zz,Pollney:2007ss,Reisswig:2009us}. In all simulations the GW amplitude is 
accurate to \emph{at least} 10\% and the phase to \emph{at least} 
1~rad over the duration of the simulation. Most of the waveforms employed 
in the construction of the analytical templates are significantly longer
(12--22 cycles) and more accurate~\cite{Hannam:2010ec}. 

We used seven sets of simulations: (1) Equal-mass binaries 
with equal, non-precessing spins $\chi_i = \pm \{0.25,0.5,0.75,0.85\}$, described 
in ~\cite{Hannam:2007wf,Hannam:2010ec}. (2) Non-precessing, equal-spin  
binaries with $q \equiv m_1/m_2 = \{2, 2.5, 3\}$ and $\chi_i = \{\pm 0.5,0.75\}$. 
(3) Nonspinning binaries with $q = \{1, 1.5, 2, 2.5, 3, 3.5, 4\}$. (4) Unequal-spin 
binaries with $q = \{2, 3\}$ and $(\chi_1, \chi_2) = (-0.75, 0.75)$. 
(5) Equal-mass, unequal-spin binaries with $\chi_i = \pm \{0.2,0.3,0.4,0.6\}$. 
(6) Equal-mass, precessing binaries with spin vectors $(0.42, 0, 0.42), (0, 0, 0)$ 
and $(0.15, 0, 0), (0, 0, 0)$. (7) Precessing $q = 3$ binary with spins $(0.75, 0, 0), (0, 0, 0)$~\cite{Schmidt:2010it}. 
Simulation sets (1)--(4) and (7) were performed with  {\tt BAM}, set (5) with 
{\tt CCATIE}, and set (6) with {\tt LLAMA}. The analytical waveform family is constructed 
employing  \emph{only} the equal-spin simulation sets (1)--(3); sets (4)--(7) were 
used to test the efficacy of our model against more general spin/mass configurations. 
Two additional waveforms were used in these tests: the Caltech-Cornell  equal-mass, 
nonspinning simulation~\cite{Scheel:2008rj}, and the RIT $q = 1.25$ precessing binary 
simulation with $|\chi_1| = 0.6, |\chi_2| = 0.4$~\cite{Campanelli:2008nk}.  

%%%%%%%%%%%%%%%%%%%%%%%%%%%%%%%%% Hybrid waveforms %%%%%%%%%%%%%%%%%%%%%%%%%%%%%%%%%%%%%
\paragraph{Constructing hybrid waveforms.---}

We produce a set of ``hybrid waveforms'' \cite{Ajith:2007qp}
by matching PN and numerical-relativity (NR) waveforms in an overlapping time interval 
$[t_1,t_2]$. These hybrids are assumed to be the target signals that we want to detect.  
For the PN waveforms we choose the ``TaylorT1''
waveforms at 3.5PN~\cite{BDEI04} phase accuracy, with spin terms up to 
2.5PN~\cite{Blanchet:2006gy,Arun:2008kb}. This is motivated by PN-NR comparisons 
of equal-mass spinning binaries, in which the accuracy of the TaylorT1 approximant
was found to be the most robust~\cite{Hannam:2007wf,Hannam:2010ec}. We include the 
3PN amplitude corrections to the dominant quadrupole mode~\cite{Blanchet:2008je} and 
the 2PN spin-dependent corrections~\cite{Arun:2008kb}, which greatly improved the 
agreement between PN and NR waveforms. For precessing waveforms, spin and angular
momenta are evolved according to~\cite{BCV2,Blanchet:2006gy}. 

We match the PN and NR waveforms by doing a least-square fit
over time- and phase shifts between the waveforms, and a
scale factor $a$ that reduces the PN-NR amplitude difference~\cite{Ajith:2007qp}. The NR waveforms are 
combined with the ``best-matched''  PN waveforms in the following way:
$\hc^{\rm hyb}(t) \equiv \, a \tau(t) \, \hc^{\rm NR}(t)+(1-\tau(t))
\, \hc^{\rm PN}(t)$, where $\hc(t)=h_+(t)-\rmi h_\times(t)$ and $\tau$ ranges 
linearly from zero to one for $t\in[t_1,t_2]$.

%%%%%%%%%%%%%%%%%%%%%%%%%%%%%%%%%%%% Templates %%%%%%%%%%%%%%%%%%%%%%%%%%%%%%%%%%%%%%%
\paragraph{Waveform templates for non-precessing binaries.---}

The analytical waveforms that we construct are written in the Fourier domain
as $h(f) \equiv {A} (f) \, e^{-\rmi\Psi (f)}$, where 
\ber
A(f) &\equiv& \mathcal{C} \fmerg^{-7/6}
\left\{ \begin{array}{ll}
f'^{-7/6} \, (1+ \sum_{i=2}^{3} \alpha_i \, v^{i})   & \textrm{if $f < \fmerg$}\\
w_m \, f'^{-2/3} \, (1+ \sum_{i=1}^{2} \epsilon_i \, v^{i}) & \textrm{if $\fmerg \leq f < \fring$}\\
w_r \, {\cal L}(f,\fring,\sigma) & \textrm{if $\fring \leq f < \fcut$,}\\
\end{array} \right. \nonumber \\
\Psi(f) &\equiv& 2 \pi f t_0 + \varphi_0 + \frac{3}{128 \,\eta\, v^5} 
\big(1 + \sum_{k=2}^{7} v^{k} \, \psi_k \big).
\label{eq:phenWaveAmpAndPhase}
\eer
Above, $f'\equiv f/\fmerg$, $v \equiv (\pi Mf)^{1/3}$,
$\epsilon_1 =  1.4547 \, \chi - 1.8897, \epsilon_2 = -1.8153 \, \chi + 1.6557$ 
(estimated from hybrid waveforms), $\mathcal{C}$ is a numerical constant whose value depends 
on the sky-location, orientation and the masses, 
$\alpha_2 = -323/224 + 451\,\eta/168$ and $\alpha_3 = (27/8 - 11\,\eta/6)\chi$
are the PN corrections to the Fourier domain amplitude of
the ($\ell=2, m=\pm2$ mode) PN waveform~\cite{Arun:2008kb}, 
$t_0$ is the time of arrival of the signal at the detector and 
$\varphi_0$ the corresponding phase, ${\cal L}(f,\fring,\sigma)$
a Lorentzian function with width $\sigma$ centered around the frequency  
$\fring$, $w_m$ and $w_r$ are normalization constants chosen so as to make 
${A}(f)$ continuous across the ``transition'' frequencies $\fring$ 
and $\fmerg$, and $\fcut$ is a convenient cutoff frequency such 
that the signal power above $\fcut$ is negligible.
The phenomenological parameters $\psi_k$ and
$\mu_k \equiv \{\fmerg, \fring, \sigma, \fcut\}$ 
are written in terms of the physical parameters of the binary as: 
\begin{equation}
\psi_k = \psi_k^0 + {\textstyle\sum\limits_{i=1}^{3}} {\textstyle\sum\limits_{j=0}^{N}} x_k^{({ij})} \eta^i \chi^j, ~~
\pi M \mu_k = \mu_k^0 + {\textstyle\sum\limits_{i=1}^{3}} {\textstyle\sum\limits_{j=0}^{N}} y_k^{({ij})} \eta^i \chi^j \,,
\label{eq:ampParams}
\end{equation}
where $N \equiv \mathrm{min}(3-i,2)$ while $x_k^{({ij})}$ and $y_k^{({ij})}$ are tabulated in
Table~\ref{tab:polCoeffsSpinAmp}.
 
We match these waveforms to 2PN accurate adiabatic inspiral waveforms in the 
test-mass ($\eta \rightarrow 0$) limit, where the phenomenological 
parameters reduce to:  
\be
\fmerg \rightarrow f^0_{\rm LSO}, ~~ \fring \rightarrow f^0_{\rm QNM}, ~~
\sigma \rightarrow f^0_{\rm QNM}/Q^0, ~~\psi_k \rightarrow \psi_k^0. 
\ee
Above, $f^0_{\rm LSO}$ and $f^0_{\rm QNM}$ are the frequencies of the
last stable orbit~\cite{BPT:1972} and the dominant
quasi-normal mode, and $Q^0$ is the ring-down quality
factor~\cite{Echevarria:1988} of a Kerr BH with mass $M$ and spin
$\chi$, while $\psi_k^0$ are the (2PN) Fourier domain phasing coefficients
of a test-particle inspiralling into the Kerr BH~\cite{Arun:2008kb}. 

The test-mass-limit waveforms suffer from two limitations:
1) we assume that the evolution of the GW phase at the merger and
ringdown is a continuation of the adiabatic inspiral phase, and 2) 
in the absence of a reliable plunge model, we approximate the amplitude of 
the plunge with $f'^{-2/3} \, (1+ \sum_{i=1}^{2} \epsilon_i \, v^{i})$.
Nevertheless, in the test-mass limit, the signal is expected to be 
dominated by the inspiral, which is guaranteed to be well-modelled by our waveforms. 
More importantly, the imposition of the appropriate test-mass limit in our
fitting procedure ensures that the waveforms are well behaved even
outside the parameter range where current NR data are available.   
Because of this, and the inclusion of the PN amplitude corrections, 
these waveforms are expected to be closer to the actual 
signals than the templates proposed in~\cite{Ajith:2007kx,Ajith:2007xh}
in the non-spinning limit. 
However, since the parameter space covered by the NR simulations is
limited, we recommend that these waveforms be used only in the regime 
$q \lesssim 10$ and $-0.85 \lesssim \chi \lesssim 0.85$. Also, these 
are meant to model only the late-inspiral, merger and ring down 
($Mf_{\mathrm{GW}} > 10^{-3}$), \ie, binaries in the mass-range 
where merger-ringdown also contribute to the SNR, apart from inspiral. 
 
%%%%%%%%%%%%%%%%%%%%%%%%%%% TABLES polynomial coefficients in spin %%%%%%%%%%%%%%%%%%%%%%%%%%%
\begin{table*}[tb]
    \begin{center}
        \begin{tabular}{ccccccccccccccc}
            \hline
            \hline
            &\vline&    Test-mass limit $(\psi^0_k)$ 	& $x^{(10)}$ 	& $x^{(11)}$ 	& $x^{(12)}$ 	& $x^{(20)}$ 	& $x^{(21)}$ 	& $x^{(30)}$ \\
\hline
$\psi_2 $ &\vline& $3715/756$ 	 & -920.9 	 & 492.1 	 &   135 	 &  6742 	 & -1053 	 & -1.34$\times 10^{4}$\\
$\psi_3 $ &\vline& $-16 \pi + 113\,\chi/3$ 	 & 1.702$\times 10^{4}$ 	 & -9566 	 & -2182 	 & -1.214$\times 10^{5}$ 	 & 2.075$\times 10^{4}$ 	 & 2.386$\times 10^{5}$\\
$\psi_4 $ &\vline&  $15293365/508032 - 405\,\chi^2/8$ & -1.254$\times 10^{5}$ 	 & 7.507$\times 10^{4}$ 	 & 1.338$\times 10^{4}$ 	 & 8.735$\times 10^{5}$ 	 & -1.657$\times 10^{5}$ 	 & -1.694$\times 10^{6}$\\
$\psi_6 $ &\vline&     0	 & -8.898$\times 10^{5}$ & 6.31$\times 10^{5}$ 	 & 5.068$\times 10^{4}$ 	 & 5.981$\times 10^{6}$ 	 & -1.415$\times 10^{6}$ 	 & -1.128$\times 10^{7}$\\
$\psi_7 $ &\vline&     0	 & 8.696$\times 10^{5}$ & -6.71$\times 10^{5}$ 	 & -3.008$\times 10^{4}$ 	 & -5.838$\times 10^{6}$ 	 & 1.514$\times 10^{6}$ 	 & 1.089$\times 10^{7}$\\
\hline
&\vline&    Test-mass limit $(\mu^0_k)$   & $y^{(10)}$	& $y^{(11)}$ 	& $y^{(12)}$ 	& $y^{(20)}$ 	& $y^{(21)}$ 	& $y^{(30)}$ \\
\hline
$\fmerg$ &\vline& $1 - 4.455(1-\chi)^{0.217} + 3.521(1-\chi)^{0.26}$ & 0.6437 	 & 0.827 	 & -0.2706 	 & -0.05822 	 & -3.935 	 & -7.092\\
$\fring$ &\vline& $[1-0.63(1-\chi)^{0.3}]/2$  & 0.1469 	 & -0.1228 	 & -0.02609 	 & -0.0249 	 & 0.1701 	 & 2.325\\
$\sigma$ &\vline&  $[1-0.63(1-\chi)^{0.3}]\, (1-\chi)^{0.45}/4$ & -0.4098 	 & -0.03523 	 & 0.1008 	 & 1.829 	 & -0.02017 	 & -2.87\\
$\fcut $  &\vline& $0.3236 + 0.04894 \,\chi + 0.01346\, \chi^2$ 	 & -0.1331 	 & -0.08172 	 & 0.1451 	 & -0.2714 	 & 0.1279 	 & 4.922\\

	        \hline
            \hline
        \end{tabular}
        \caption{Phenomenological parameters describing the analytical waveforms. 
        In test-mass limit, they reduce to the 
        appropriate quantities given by perturbative calculations~\cite{Arun:2008kb,
        BPT:1972,Echevarria:1988}. The test-mass limit of $\fmerg$ is 
        a fit to the frequency of the last stable orbit~\cite{BPT:1972}.}
        \label{tab:polCoeffsSpinAmp}
    \end{center}
\end{table*}
%%%%%%%%%%%%%%%%%%%%%%%%%%%%%%%%%%%%%%%%%%%%%%%%%%%%%%%%%%%%%%%%%%%%%%%%%%%%%%%%

%%%%%%%%%%%%%%%%%%%%%%%%%%%%%%%%%%%%%%%%%%%%%%%%%%%%%%%%%%%%%%%%%%%%%%%%%%%%%%%%
\begin{figure}[tb]
\centering
\includegraphics[width=3.1in]{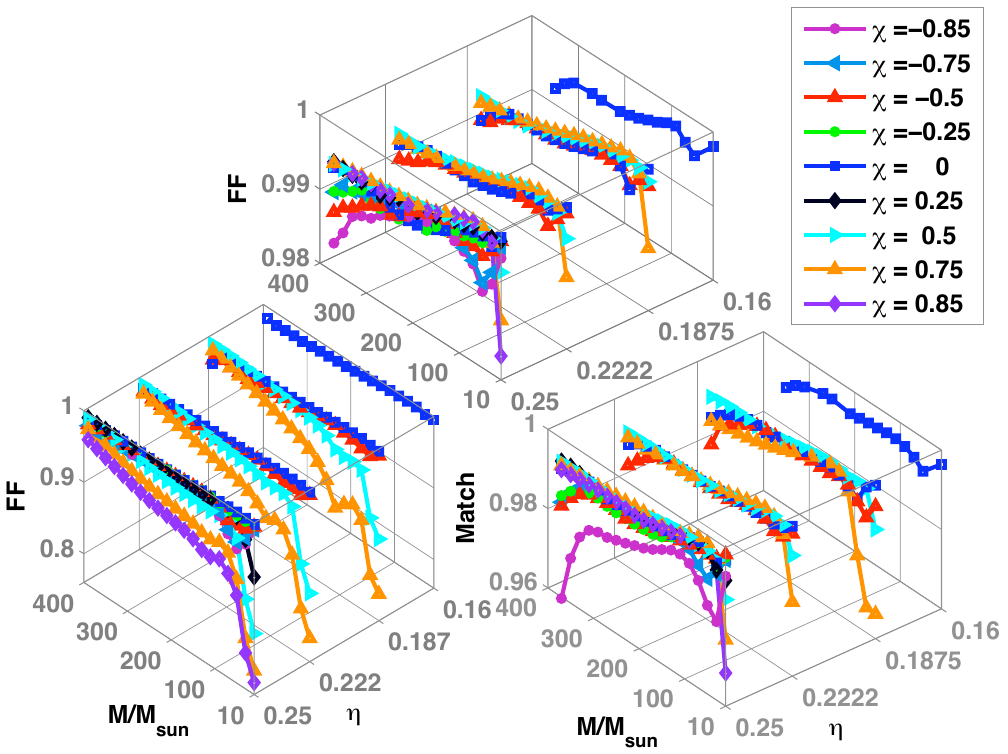}
\caption{\emph{Top and right plots}: Match and FF of our IMR templates 
with equal-spin hybrid waveforms constructed from simulation 
sets (1)--(3). \emph{Bottom left}: FF of \emph{non-spinning} IMR templates proposed 
in~\cite{Ajith:2007kx,Ajith:2007xh} with the same hybrids.} 
\label{fig:FFAndFaithfulSpinLIGO}
\end{figure}

\begin{figure}[tb]
\centering
\includegraphics[width=3.1in]{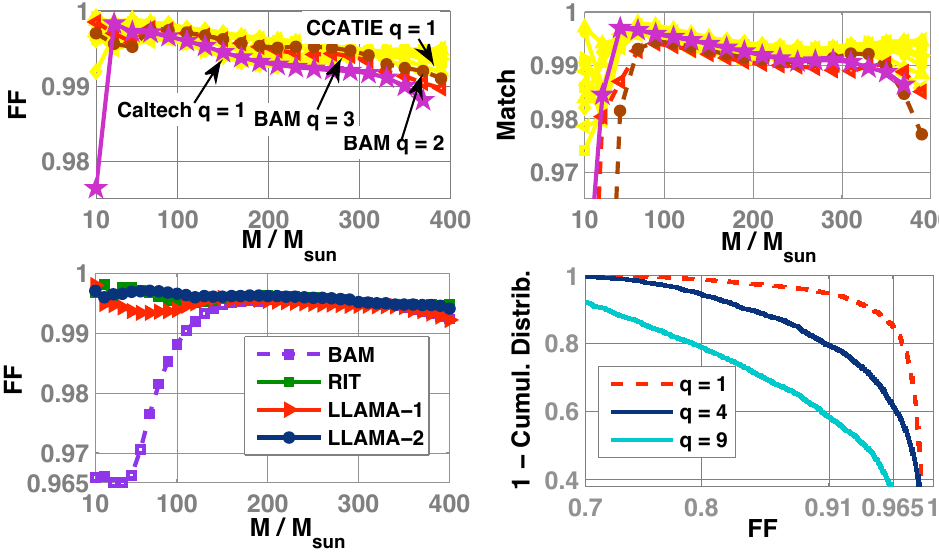}
\caption{Top panel: Match and FF of our templates with 
unequal-spin hybrid waveforms constructed from simulation sets (iv) and (v),
and the Caltech-Cornell non-spinning simulation. Bottom left: FF with 
precessing hybrids constructed from sets (vi) and (vii),
and the RIT simulation. Bottom right: Fraction of precessing 
PN waveforms ($M=10 M_\odot$) producing fitting factor FF with the IMR templates 
--- 85\% (62\%) 37\% of the binaries with $q = 1\, (4) \,9$ 
produce FF $> 0.965$.}
\label{fig:FFAndFaithfulSpinLIGOUneqSpin}
\end{figure}
%%%%%%%%%%%%%%%%%%%%%%%%%%%%%%%%%%%%%%%%%%%%%%%%%%%%%%%%%%%%%%%%%%%%%%%%%%%%%%

We have examined the ``faithfulness''~\cite{DIS98} of the new templates in reproducing the hybrid 
waveforms by computing the \emph{match} (noise-weighted inner product) 
with the hybrids. Loss of the SNR due to the ``mismatch'' between 
the template and the true signal is determined by the match maximized over the 
whole template bank -- called \emph{fitting factor} (FF). The standard criteria for templates 
used in searches is that FF $>$ 0.965, which corresponds to a loss of no 
more than 10\% of signals.

Match and FF of the analytical waveforms with the equal- (unequal-) spin hybrid waveforms 
are plotted in Fig.~\ref{fig:FFAndFaithfulSpinLIGO} (Fig.~\ref{fig:FFAndFaithfulSpinLIGOUneqSpin}), 
using the Initial LIGO design noise spectrum~\cite{InitLIGOSRD}. Note that the analytical 
waveform family is constructed employing \emph{only} the equal-spin hybrid waveforms 
(Fig.~\ref{fig:FFAndFaithfulSpinLIGO}). The PN--NR matching region used to construct the
unequal-spin hybrids (Fig.~\ref{fig:FFAndFaithfulSpinLIGOUneqSpin}) are also different from 
that used for equal-spin hybrids. These figures demonstrate the efficacy of the analytical 
templates in reproducing the target waveforms -- templates are ``faithful'' (match $>$ 0.965) 
\emph{either} when the masses \emph{or} the spins are equal, while they are \emph{always} 
``effectual''~\cite{DIS98} in detection (FF $>$ 0.965). In contrast, the bottom 
left plot of Fig.~\ref{fig:FFAndFaithfulSpinLIGO} shows the FF of the non-spinning IMR template 
family proposed in~\cite{Ajith:2007kx,Ajith:2007xh} with the equal-spin hybrids. 
FFs as low as 0.8 suggest that up to 50\% of binaries may go undetected if nonspinning 
IMR templates are employed to search for binaries with high (aligned) spins. 

The bottom left plot of Fig.~\ref{fig:FFAndFaithfulSpinLIGOUneqSpin} shows the FF and match 
of the template family with four \emph{precessing} hybrids. The high FFs are
indicative of the effectualness of the templates in detecting precessing binaries. 
Since presently not enough NR simulations are available to make a quantitative
statement, and since we expect the effect of precession will be predominant in the
case of lower mass binaries (when large number of cycles are present in the detector band), 
we might be able to acquire some useful indication by studying precessing PN waveforms. 
We performed a Monte-Carlo simulation where we generate precessing
``restricted'' PN waveforms with $M = 10 M_\odot$, $q = \{1, 4, 9\}$, 
uniformly distributed spin magnitudes in the interval $[0, 0.98]$ and isotropically distributed 
spin angles, and compute the FF with the templates proposed in this paper. The inclination of 
the binary's total angular momentum with the line of sight from the observer is also randomly 
chosen from $[0, \pi]$.  The bottom right plot of Fig.~\ref{fig:FFAndFaithfulSpinLIGOUneqSpin} 
shows the cumulative distribution of the FF, strongly indicating the effectualness 
of the templates in detecting precessing binaries in the comparable-mass regime.
These results indicate that a search employing non-precessing  templates described 
by a single spin parameter might be an attractive and feasible way of searching 
for generic spinning binaries~\footnote{These results hold for Enhanced LIGO also.}. 

\begin{figure}[tb]
\centering
\includegraphics[width=3.0in]{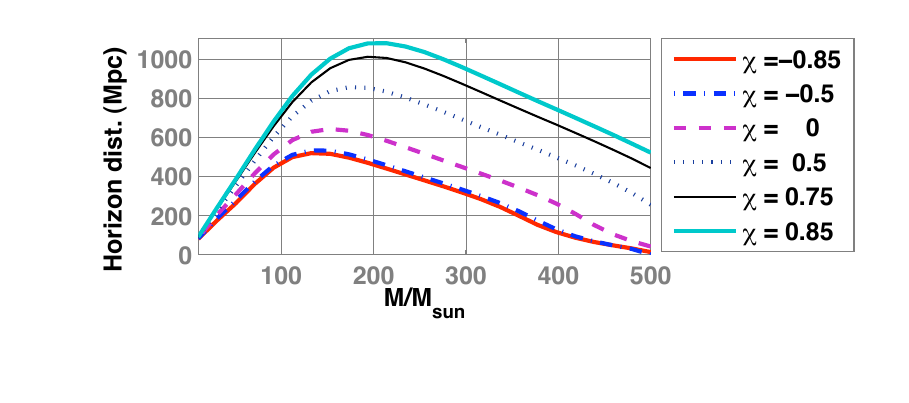}
\caption{Distance to optimally oriented equal-mass binaries with 
spin $\chi$ producing SNR 8 in Initial LIGO.}
\label{fig:HorDistSpinLIGO}
\end{figure}

Distance to optimally oriented BBHs producing SNR of 8 in Initial LIGO 
is shown in Fig.~\ref{fig:HorDistSpinLIGO}, which demonstrates the dramatic 
effect of spin for detection of high-mass binaries; if most BBHs are
highly spinning, then LIGO will be able to detect coalescences up to 1Gpc, thus
increasing the event rates as much as five times compared to predictions based 
on models of nonspinning binaries. For Advanced LIGO, the distance reach is as 
high as 20 Gpc.

%%%%%%%%%%%%%%%%%%%%%%%%%%%%%%%%%%%%%%%%%%%%%%%%%%%%%%%%%%%%%%%%%%%%%%%%%%%%%%%%
%%%%%%%%%%%%%%%%%%%%%%%%%%%%%%%%%%%%%%%%%%%%%%%%%%%%%%%%%%%%%%%%%%%%%%%%%%%%%%

\paragraph{Conclusions.---}
We combine state-of-the-art results from analytical- and numerical relativity to 
construct a family of analytical IMR waveforms for BBHs with 
non-precessing spins. These waveforms are also able to 
detect a significant fraction of the precessing binaries in the comparable-mass
regime, with spins represented by a \emph{single} parameter. This will considerably simplify the use of our 
waveforms in GW searches in the near future, and will accelerate the incorporation 
of NR results into the current effort for the first detection of GWs. There are 
many other immediate applications of our waveforms: injections into detector data 
will help to put more realistic upper limits on the rate of BBH coalescences, and 
to compare the different algorithms employed in the search for BBHs, while employing 
these in population-synthesis studies will provide more accurate coalescence rates 
observable by the current and future detectors. Our method can readily be generalized 
to incorporate non-quadrupole harmonics, larger portions of the BBH parameter space 
and further information from analytical approximation methods or numerical simulations.

\acknowledgments 
SH was supported by VESF and EGO, DAAD grant D/07/13385  and 
Spanish Ministry of Science grant FPA-2007-60220. MH
was supported by FWF grant M1178-N16 and STFC grant ST/H008438/1. 
PA and YC were supported by NSF grants PHY-0653653, PHY-0601459,  
PHY-0956189 and David and Barbara Groce Fund. BB is supported by DFG 
grant  SFB/Transregio~7 ``GW Astronomy'', BB and 
DM by DLR, and LS by DAAD grant A/06/12630. We thank AEI, FSU Jena, 
LRZ, ICHEC, VSC, CESGA and BSC-CNS for computational resources, 
and K.\,G.\,Arun, B.\,Sathyaprakash, G.\,Faye and R.\,O'Shaughnessy for  
discussions.

%%%%%%%%%%%%%%%%%%%%%%%%%%%%%%%%%%%% References %%%%%%%%%%%%%%%%%%%%%%%%%%%%%%%%%%%%%%%%
\bibliographystyle{apsrev-nourl}
\bibliography{SpinTempl}

\end{document}